\documentstyle[12pt,openbib]{article}
\begin{document}
\title{\bf
Observed Supersymmetry and Parity Doubling of Excited Hadrons from a
Simple Spectrum-Generating Algebra }
\baselineskip .4cm
\author{Jishnu Dey\thanks{Supported in part by DST, Govt. of
India and by FAPESP of S\~ao Paulo, Brasil}
\\ Instituto de F\'\i sica Te\'orica,  UNESP, Rua Pamplona 145 \\
S\~ao Paulo, 01405-900, SP, Brasil \\
\\ and \\
Mira Dey\thanks{Supported in part by DST, Govt. of India and
on leave from Department of Physics, Lady Brabourne College,
Calcutta 700017 India } \\ ICTP, P.O. Box 386, Trieste, Italy
34100,\\ }

\date{\today }
\maketitle
{\it Abstract} : We find remarkable agreement with the observed
excitations of hadrons with a simple three parameter mass
relation of the SU(3) subgroup of the underlying U(15/30) graded
Lie group. The baryons are the appropriate supersymmetric
partners of the mesons. An interesting feature, which is a focus
of current interest, is that the baryons  and isobars show
parity doubling. Significantly, the ground state baryons and
mesons have no place in the fit, so that the parity doubling is
indicated only when excitation energy is available. This has
correspondence with the parity doubling seen in recent lattice
calculations, when thermal excitation is present.

The agreement with experiment is comparable to the
semi-relativistic quark model which fits with many
force parameters and where the observed parity doubling is
accidental. The gross splitting of the levels is the same for
the strange and the non-strange sectors, suggesting flavour
independence.

\newpage
\baselineskip .7cm
\noindent {\bf 1. Introduction.}

\vspace{.5cm}

Dynamical groups and spectrum generating algebras are much used
in the current literature for classifying states of a composite
system in terms of underlying group symmetry of the Hamiltonian.
The pioneers in this field were Barut \cite{bar1}, Barut and
Bohm \cite{bar2} and Dothan, Gell Mann and Ne'emna\cite{dot}.
The subject has been reveiwed in detail recently in two volumes
edited by Bohm, Ne'eman and Barut \cite{boh}.

Lattice calculations and other QCD motivated  models
exist and they indicate that the phase transition to free quarks
and gluons is a weak first order or second order one. So there
is no  dissolution  of  the hadrons,  but  rather one sees
presence of parity doublets \cite{dk}.  This was soon confirmed
by Gottleib et al \cite{gott1}

In this connection we must stress the early work of Barut
\cite{bar1} where he had looked at parity doubled states in the
conformal O(4,2) model. The conformal model has been revived in
the version of string  theory  given by Kutasov  and  Seiberg
\cite{kut} and leads to a rich phenomenology as shown by the
Freund and Rosner \cite{fr},  Dey, Dey  and  Tomio
\cite{ddt}, Cudell and Dienes \cite{cud} and Mustafa et
al.\cite{mus}.  This is based on old wisdom of the  Regge
model giving  same  trajectory  for  mesons  and  baryons.
According to Kutasov and Seiberg, the  appearance  of  the
destabilizing tachyons in a string theory severely constrains
the difference  of the densities of bosons and fermions in that
theory. Their  result shows that  tachyon  elimination  does
not  require  full-fledged supersymmetry. Cancellation between
the boson and  fermion  states is all that is needed. It turns
out that  though  the  density  of states of mesons and baryons
each rises exponentially with energy, their difference rises
only like a low power  of  energy. The present model uses a much
simpler compact group structure and our only motivation is to
prompt more calculations based on the phenomenological success
of the model proposed. We retain the same common feature of the
two very different calculations \cite{bar1} and \cite{dk},
namely  that the ground state does not show parity doubling but
the excited states of the baryons do.

\vskip 1.5cm
\noindent {\bf 2. The Model}
\vskip .5cm

Looking at the baryon spectrum we see that the lowest states, i.e.
the nucleon octet ($N, \Lambda , \Sigma, \Xi $), have no odd
parity partners. But when some excitation energy is available,
we encounter parity doubling. It is no surprising that one can
get this at finite T, in lattice as we have already discussed.
The Laplace transform of the finite temperature partition
function in fact gives the excitation spectrum at zero
temperature \cite{ddt}.

Let us take some examples, one can think of $N(1535)$, $\Lambda
(1405)$, $1/2^{-}$ states $N(1440)$, $\Lambda (1600)$, $1/2^{+}$
; the $N(1675)$, $5/2^{-}$ is almost degenerate with the
$N(1680)$, $5/2^{+}$ state  etc.  Chiral symmetry is realized at
such high excitation within 5-10 \%. In the non-relativistic and
the semi-relativistic models, the occurrence of parity doubled
states is contrived : the perturbative hyperfine interaction is
adjusted to bring down the even parity to match with the odd
levels, sometimes invoking multi-shell configurations \cite{cap},
sometimes deformation \cite{mur}.

In an earlier paper \cite{dey}, we had fitted mesons and baryons
in a simple model using the supersymmetric graded Lie group
U(15/30). We add the odd parity baryons into our scheme, in view
of the present interest. In the meson sector the interest is of
a different nature. Recently many more new mesons sates have been
discovered and these do not all fit into any single model. We
discuss our fit in terms of the new experimental data.

Our model is based on excitations of bosons and fermions in the
s, d and g shells of some effective potential, in terms of a
U(15/30) graded Lie group. The reduction of this group into
simpler structures, in particular to the SU(3) scheme, so well
known in Nuclear Physics \cite{ell} have been worked out
\cite{kot}, \cite{yu}. In \cite{dey} we had given a mass formula
which fitted more than 60 mesons and baryons using the
classification given by Yu. We now add more than 40 new states.

The scheme we follow is extremely simple with states belonging
to a representation where the total number of particles is
three. This implies that the mesons are different from other
models, here they are two fermion, one boson states : $f\bar f b$.
The alternative is to use the meson as a two fermion and the
baryon as a quark-diquark state, and assume diquarks are similar
to antiquarks. But in  view of the complexity of the recently
discovered excited mesons \cite{kon}, we have preferred the
present model. These fermions and bosons are colour-dressed
quasiparticles, they could be quasi-quarks and quasi-gluons or
other configurations.

The meson and baryon states are then classified accroding to the
partition $[\nu ]$ and the SU(3) representation $(\lambda , \mu )$.
The SU(3) Casimir operators are $L ( L + 1 )$ and :

\begin{equation}
C(\lambda , \mu ) =\lambda ^{2} + \mu ^{2} + \lambda \mu + 3
(\lambda + \mu )
\label{eq:cas}
\end{equation}

The spin zero mesons may belong to partition $\nu = [3], [21]$.
Spin one mesons and octet baryons (spin $1/2$) can belong to the
latter partition. This and the allowed values of $(\lambda , \mu
)$ are given in Table 1. The isobars and the J = 1 mesons can be
placed in the partiton $\nu = [111]$. The allowed $(\lambda ,
\mu )$ are also given in Table 1.

The hadron mass $M$  is then given by

\begin{equation}
M = 2700 - 9 C(\lambda , \mu ) + 8 S(S + 1) + \alpha L(L+1)
\label{eq:m}
\end{equation}

In a simple model like ours it is not justified to try for
accuracy, particularly in view of the typical width of 50 MeV
for the states. However one can associate $\alpha $ with the
inverse of a moment of inertia parameter, and since the masses
and the radii vary for different bandheads, given by a set
$(\lambda , \mu )$, we have allowed $\alpha $ to vary a little.
\vskip .5cm
\noindent {\bf 3. Results and discussion}
\vskip .5cm

We now turn to baryons which are given in Tables 2 to 5.
Unlike \cite{cap} all levels are fitted and the fit is generally
similar. The Roper resonance N(1440) and the
N(1710) are fitted very well (Table 2). The placement of 1710 in
the second band is consistent with its very different properties
: it has a gamma decay width consistent with zero and ten times
less than the Roper. It also has a very large two pion decay
rate and very little $N\pi $ unlike the Roper. Altogether 18
positiv parity baryons are fitted in Table 2. The quality of the
fit is very good, the difference from experiment being 5.7\%. All
the well-established $\Lambda $-s  and $\Sigma $-s
 are fitted along with three two-star $\Sigma $-s. It is
interesting to note that in Table 2, and indeed also in the
subsequent tables, the strange baryons fit along side nucelon
excitations. On the other hand recall, that the splitting
between the ground state multiplet is dependent on the mass of
particular quark flavours, for example in the recent Shuryak and
Rosner model \cite{sr}. Thus one might say that the gross
spacing of hadron excitations is independent of flavour, but the
hyperfine splitting between the states, particularly the ground
states, is known to depend on the masses of its quark content.

	We include the odd parity baryons in our fit (Table 3)
and suggest some parity doublets. In \cite{ir} the same is done,
but there is a very important point of difference between
Iachello's work and ours. His model and the variant suggested by
Robson \cite{ir} depend essentially on geometrically symmetric
configurations. It is difficult to envisage why the ground
states may be left out of such geometric symmetries. Our model,
on the other hand uses chiral symmetry restoration, which happens
only for excited configurations. The phenomenology may find
justification only in an effective potential based on non-compact
conformal group structure as indicated in Barut's work
\cite{bar1}. Notice that the spectrum given by him has parity
doubling,  not in the ground state, but starting from the first
excited state.

	We have placed some of the odd parity baryons as L = 0
excitations of odd-parity quasiparticles in Table 3. This gives
much better fit and fits in with the fact that $\Lambda (1405)$
and the $N(1535)$ are different from the $1520$ $\Lambda $ or
$N$.  This is supported by the rather unusual 100 \% $\Sigma
\;\pi $ decay of the first one and the 45-55\% $N\eta $ decay of
the second one. The reader who does not like such exotic
quasiparticles may group them together with the $(L = 1)$ $1520$
  $\Lambda $ and $N$.

	The number of odd parity baryons fitted in Table 3 is
24, the fit is comparable to that of Table 2. In fact using all
the 42 states the fit improves  marginally to 5.03\%.

	We fit 8 positive and 6 negative parity baryons in
Tables 4, 5 for which $S = 3/2$ in eq. (2) and $\alpha $ is
taken to be $30 $ $MeV$. The fit to experiment is 2.93\%. The
experimental states consist of 9 well establsihed and 5 states
with (**) status.

	In Table 6 we have placed 26 mesons with S = 0. The
experimental states are shown for comparison, their
summary - table - status are indicated. The partition $[\nu = 3]$
and $[\nu = 21]$ show a degeneracy for $C(\lambda , \mu ) = 114
$.  They mix and the bracketed numbers 1574 and 1774 would
result from a $100$ $MeV$ off-diagonal interaction between them.
The same situation is found for $C(\lambda , \mu ) = 90 $ and
there the off-diagonal interaction is suggested to be $200$
$MeV$. The fit to experiment averages to 4.75 \%.

	In \cite{go} there are six $f_2$ which fit with five of
the six observed confirmed states, left out is the $2340$ $MeV$
state. In our model all eight observed $f_2$ states are fitted
in a natural way, while four  $f_2$ states at $1430$,
 $1640$, $1810$, $2150$ and $2175$ are left out of the summary tables,
although the $1810$ has been
seen by three different groups. The $1565$ has also been
confirmed now but may be a diquark-antidiquark state as already
mentioned \cite{kon}.

	In Table 8 the $\rho $, $\omega $, $\phi $, and $K^* $
are compared with the S= 1 mesons. There are 20 states and the
fit to experiment averages to 4.15 \%. Altogether we fit 102
hadrons and the fit is about 5 \%.

	In summary we have fitted 56 baryons, 26 spin zero
mesons and 20 spin one mesons with two fixed parameters and
another which varies a little from 30 MeV to 45 MeV. In an
effective potential such a sdg-structure is natural. The SU(3)
scheme diagonalizes the first two leading terms of the expansion
of the effective potential about a local minimum, namely the
oscillator term and the quadrupole term. Also experience with
nuclear physics shows its usefulnessness in effective potential
models. Finally one can look for decay systematics, but for this
one has to impose a model ground state structure, since the
ground states are left out of the scheme. A conformal model
which incorporates the ground state in  consistent way may be
able to achieve this.

\vskip .3cm
\noindent {\bf Acknowledgements.}
\vskip .3cm

Mira Dey wishes to thank the ICTP for supporting the visit as an associate
(Summer, 1993). J. Dey
acknowledges support from FAPESP of S. P., Brasil. The  Deys also acknowledge
partial support from DST (Govt. of India, grant no. SP/S2/K04/92).

\newpage

\begin{table}
\centering
\caption{The partition [$\nu $], SU(3) Young representation
$(\lambda , \mu )$, the fermion number $N_{F}$ and spin S given
for $N \equiv N_{F} + N_{B} = 3$}

\vspace{.5 cm}
\begin{tabular}{c|llllllll}
\hline
  $[\nu ]$ &  &  &  & $(\lambda , \mu )$ &  &  &  &$(N_{F},S)$   \cr
\hline
 $[3]$& (12,0) & (8,2) & (6,3) & (6,0) & (4,4) & (3,3) & ....&(2,0) \cr
 \hline
 $[21]$& (10,1) & (8,2) & (6,3) & (7,1) & (6,3) & (6,0) & ....&(2,0) \cr
 & & & & & & & &(2,1) \cr
 & & & & & & & &(3,1/2) \cr
\hline
 $[111]$& (9,0) & (6,3) & (6,2) & (3,3) & (3,0) & (2,5) & ....&(2,1) \cr
 & & & & & & & &(3,3/2) \cr
\hline
\end{tabular}
\end{table}

\newpage

\begin{table}
\centering
\caption{Even Parity baryons $\alpha = 40\; MeV$. All energies in $MeV$}

\begin{tabular}{cccccccc}
\\
\hline
\multicolumn{1}{c}{\ \ Band} &
\multicolumn{1}{c}{\ \ $C(\lambda , \mu )$} &
\multicolumn{1}{c}{ \ \ $L$ } &
\multicolumn{1}{c}{IBFM }&
\multicolumn{1}{c}{$J^{P}$}&
\multicolumn{1}{c}{$Experiment$}&
\multicolumn{1}{c}{RF. [12]}
 \\
\hline
 $1$ & 144 & 0 & 1410 & $1/2^{+}$ & $N(1440) : (1430-1470)$ & 1540 \\
      &     &  &      &           & $\Lambda (1600) : (1560-1700)$ & 1680 \\
\vspace{.1 cm}
      &     & 2 & 1650 & $5/2^{+}$ & $N (1680) : (1675-1690)$ & 1770 \\
      &     &  &       &           & $\Lambda (1820) : (1815-1825)$ & 1890 \\
      &     & 2 &      & $3/2^{+}$ & $N (1720) : (1650-1750)$ & 1795 \\
\vspace{.1 cm}
      &     & 4 & 2210 & $9/2^{+}$ & $N (2210) : (2180-2310)$ &   -   \\
      &     &   &      &           & $\Lambda (2350) : (2340-2370)$ & - \\
      &     & 4 &      & $7/2^{+}$ & $\Sigma (2030) : (2025-2040)$ & 2080 \\
      &     &   &      &           & $N (1990) : (**)$ & 2000 \\
\vspace{.1 cm}
      &     & 6 & 3090 & $13/2^{+}$ & $ N(2700) : (**)$ & - \\
\hline
 $2$ & 114 & 0 & 1680 & $1/2^{+}$ & $N(1710) : (1680-1740)$ & 1770 \\
      &     &  &      &           & $\Lambda (1810) : (1750-1850)$ & 1830 \\
      &     &  &      &           & $\Sigma (1660) : (1630-1690)$ & 1720 \\
\vspace{.1 cm}
      &     & 2 & 1920 & $3/2^{+}$ & $\Lambda (1890) : (1850-1910)$ & 1900 \\
      &     &   &      &           & $\Sigma (2080) : (**)$ & 2010   \\
      &     & 2 &      & $5/2^{+}$ & $\Lambda (2110) : (2090-2140)$ & 2035\\
      &     & 4 &      &           & $\Sigma (1915) : (1900-1935)$ & 1920 \\
      &     &   &      &           & $N (2000) : (**)$ & 1995 \\
\hline
\end{tabular}
\end{table}

\begin{table}
\centering
\caption{Odd Parity baryons $\alpha = 40\; MeV$. All energies in $MeV$}

\begin{tabular}{cccccccc}
\\
\hline
\multicolumn{1}{c}{\ \ Band} &
\multicolumn{1}{c}{\ \ $C(\lambda , \mu )$} &
\multicolumn{1}{c}{ \ \ $L$ } &
\multicolumn{1}{c}{IBFM }&
\multicolumn{1}{c}{$J^{P}$}&
\multicolumn{1}{c}{$Experiment$}&
\multicolumn{1}{c}{RF. [12]}
 \\
\hline
 $1$ & 144 & 0 & 1410 & $1/2^{-}$ & $N(1535) : (1520-1555)$ & 1460 \\
      &     &  &      &           & $\Lambda (1405) : (1407 \pm 4)$ & 1550 \\
\vspace{.1 cm}
      &     & 2 & 1650 & $5/2^{-}$ & $N (1675) : (1670-1685)$ & 1630 \\
\vspace{.1 cm}
      &     & 4 & 2210 & $9/2^{-}$ & $N (2250) : (2170-2310)$ &   -   \\
\vspace{.1 cm}
      &     & 1 & 1490 & $3/2^{-}$ & $\Lambda (1520) : (1519.5\pm 1)$ &1545\\
      &     &   &      &            & $N (1520) : (1510-1530)$ & 1495\\
      &     &   &      &            & $\Sigma (1580) : ( ** )$ &1655\\
      &     & 1 &      & $1/2^{-}$ & $\Sigma (1620) : ( ** )$ &1630\\
\vspace{.1 cm}
      &     & 3 & 1890 & $5/2^{-}$ & $\Sigma (1775) : (1770-1780)$ & 1755 \\
      &     &   &      &           & $\Lambda (1830) : (1810-1830)$ & 1775 \\
      &     & 3 &      & $7/2^{-}$ & $\Lambda (2100) : (2090-2110)$ & 2150 \\
\vspace{.1 cm}
      &     & 5 & 2610 & $11/2^{-}$ & $N (2600) : (2550-2750)$ & - \\
\hline
 $2$  & 114  & 0 & 1680 & $1/2^{-}$ & $N(1650) : (1640-1680)$ & 1535 \\
      &      &   &      &           & $\Sigma (1750) : (1730-1800)$ & 1675 \\
      &      &   &      &           & $\Lambda (1670) : (1660-1680)$ & 1615 \\
\vspace{.1 cm}
      &     & 2 & 1920 & $3/2^{-}$ & $\Sigma (1940) : (1900-1930)$ & 1750 \\
      &     &   &      &           & $N(2050) : ( ** )$ & 1960   \\
      &     & 2 &      & $5/2^{-}$ & $N (2200) : ( ** )$ & 2080\\
\vspace{.1 cm}
      &     & 1 & 1760 & $3/2^{-}$ & $\Sigma (1670) : (1665-1685)$ & 1755 \\
      &     &   &      &           & $N (1700) : (1650-1750)$ & 1625 \\
      &     &   &      &           & $\Lambda (1690) : (1685-1695)$ & 1645 \\
\hline
 $3$  & 90  & 0 & 1896 & $1/2^{-}$ & $\Lambda (1800) : (1720-1850)$ & 1675 \\
\vspace{.1 cm}
      &     & 1 & 1976 & $3/2^{-}$ & $\Sigma (1940) : (1900-1950)$ & 1750 \\
\end{tabular}
\end{table}

\begin{table}
\centering
\caption{Even Parity isobars $\alpha = 30\; MeV$. All energies in $MeV$}

\begin{tabular}{cccccccc}
\\
\hline
\multicolumn{1}{c}{\ \ Band} &
\multicolumn{1}{c}{\ \ $C(\lambda , \mu )$} &
\multicolumn{1}{c}{ \ \ $L$ } &
\multicolumn{1}{c}{IBFM }&
\multicolumn{1}{c}{$J^{P}$}&
\multicolumn{1}{c}{$Experiment$}&
\multicolumn{1}{c}{RF. [12]}
 \\
\hline
 $1$ & 108 & 0 & 1758&$3/2^{+}$ &$\Delta(1600):(1550-1700)$& 1540 \\
\vspace{.1 cm}
      &     & 2 & 1938 & $7/2^{+}$ & $\Delta (1950) : (1940-1960)$ & 1770 \\
      &     &   &      & $5/2^{+}$ & $\Delta (1905) : (1870-1920)$ & 1890 \\
      &     &   &      & $3/2^{+}$ & $\Delta (1920) : (1900-1970)$ & 1795 \\
      &     &   &      & $1/2^{+}$ & $\Delta (1910) : (1870-1920)$ & 1875 \\
\vspace{.1 cm}
      &     & 4 & 2358 & $11/2^{+}$ & $\Delta (2420) : (2300-2400)$ & - \\
      &     &   &      & $9/2^{+}$ & $\Delta (2300) : ( ** )$ & - \\
\vspace{.1 cm}
      &     & 6 & 3018 & $15/2^{+}$ & $\Delta (2950) : ( ** )$ & -  \\
\hline
\end{tabular}
\end{table}

\begin{table}
\centering
\caption{Odd Parity isobars $\alpha = 30\; MeV$. All energies in $MeV$}

\begin{tabular}{cccccccc}
\\
\hline
\multicolumn{1}{c}{\ \ Band} &
\multicolumn{1}{c}{\ \ $C(\lambda , \mu )$} &
\multicolumn{1}{c}{ \ \ $L$ } &
\multicolumn{1}{c}{IBFM }&
\multicolumn{1}{c}{$J^{P}$}&
\multicolumn{1}{c}{$Experiment$}&
\multicolumn{1}{c}{RF. \cite{cap}}
 \\
\hline
 $1$  & 108 & 0 & 1758 & $3/2^{-}$ & $\Delta(1700) : (1670-1770)$ & 1620 \\
\vspace{.1 cm}
      &     & 2 & 1938 & $1/2^{-}$ & $\Delta (1900) : (1850-1950)$ & 1770 \\
      &     &   &      & $5/2^{-}$ & $\Delta (1930) : (1920-1970)$ & 1890 \\
\vspace{.1 cm}
      &     & 4 & 2358 & $9/2^{-}$ & $\Delta (2400) : (**)$ & - \\
\vspace{.1 cm}
      &     & 1 & 1818 & $1/2^{-}$ & $\Delta (1620) : (1615-1675)$ & 1555\\
\vspace{.1 cm}
      &     & 5 & 2718 & $13/2^{-}$ & $\Delta (2750) : ( ** )$ & -  \\
\hline
\end{tabular}
\end{table}

\begin{table}
\centering
\caption{The f, a and $\pi $ families, $\alpha = 35\; MeV$. All
energies in $MeV$}

\begin{tabular}{cccccccc}
\\
\hline
\multicolumn{1}{c}{Band} &
\multicolumn{1}{c}{$[\nu ]$} &
\multicolumn{1}{c}{$C(\lambda , \mu )$} &
\multicolumn{1}{c}{ $L$ } &
\multicolumn{1}{c}{IBFM }&
\multicolumn{1}{c}{$I^{G}(J^{P})$}&
\multicolumn{1}{c}{$Expt.(with\; summary\; T. \;status)$}&
\multicolumn{1}{c}{RF. \cite{go}}
 \\
\hline
\vspace{.1 cm}
$1$ & 3  & 180 & 0 & 1080 & $0^{+}(0^+)$ & $f_0(975) : (included)$ & 1090 \\
    &    &     &   &      & $1^{-}(0^+)$ & $a_0(980) : (included)$ & 1090 \\
\vspace{.1 cm}
    &    &     & 2 & 1290 & $0^{+}(2^+)$ & $f_2(1270) : (included)$ & 1280 \\
    &    &     &   &      & $1^{-}(2^+)$ & $a_2(1320) : (included)$ & 1310 \\
\vspace{.1 cm}
    &    &     & 4 & 1780 & $0^{+}(4^+)$ & $f_4(2050) : (included)$ & 2010 \\
    &    &     &   &      & $1^{-}(4^+)$ & $a_4(2040) : (omitted )$ & 2010 \\
\vspace{.1 cm}
    &    &     & 6 & 2550 & $0^{+}(6^+)$ & $f_6(2510) : (omitted )$ &   -  \\
    &    &     &   &      & $1^{-}(6^+)$ & $a_6(2450) : (omitted )$ &   -  \\
\hline
\vspace{.1 cm}
$2$ & 21  & 144 & 0 & 1404 & $0^{+}(0^+)$ & $f_0(1400) : (included)$ & 1360 \\
    &    &     &   &      & $1^{-}(0^-)$ & $\pi (1300) : (included)$ & 1300 \\
    &    &     &   &      & $1^{-}(0^+)$ & $a_0(1320) : (omitted)$ & 1780 \\
\vspace{.1 cm}
    &    &     & 2 & 1614 & $0^{+}(2^+)$ & $f_2(1525) : (included)$ & 1530 \\
    &    &     &   &      & $1^{-}(2^-)$ & $\pi 2(1670) : (included)$ & 1620 \\
\vspace{.1 cm}
    &    &     & 4 & 1780 & $0^{+}(4^+)$ & $f_4(2220) : (omitted)$ & 2200 \\
\hline
\vspace{.1 cm}
$3$ & 3  & 114 & 0 & 1674 & $0^{+}(0^+)$ & $f_0(1525) : (omitted )$ & - \\
    &    &     &   &(1574)&              &                        &  \\
\vspace{.1 cm}
    &    &     & 2 & 1884 & $0^{+}(2^+)$ & $f_2(1810) : (omitted)$ & 1820 \\
    &    &     &   &(1784)&              &                        &  \\
\vspace{.1 cm}
    &    &     & 4 & 2374 & $0^{+}(4^+)$ & $f_4(2300) : (omitted)$ & - \\
    &    &     &   &(2274)&              &                        &  \\
\vspace{.1 cm}
$4$ & 21  & 114 & 0 & 1674 & $0^{+}(0^+)$ & $f_0(1710) : (included )$ & 1780 \\
    &    &    &   &(1774)& $1^{-}(0^-)$ & $\pi (1770) : (omitted )$ & 1880 \\
\vspace{.1 cm}
    &    &     & 2 & 1884 & $0^{+}(2^+)$ & $f_2(1810) : (omitted )$ & 2040 \\
    &    &    &   &(1984)&$1^{-}(2^-)$ & $\pi 2(2100): (omitted )$ & 2130  \\
\hline
\vspace{.1 cm}
\end{tabular}
\end{table}

\begin{table}
\centering
\caption{Continuation of the f, a and $\pi $ families, $\alpha =
35\; MeV$ as in Table 6. All energies in $MeV$}
\begin{tabular}{cccccccc}
\\
\hline
\multicolumn{1}{c}{Band} &
\multicolumn{1}{c}{$[\nu ]$} &
\multicolumn{1}{c}{$C(\lambda , \mu )$} &
\multicolumn{1}{c}{ $L$ } &
\multicolumn{1}{c}{IBFM }&
\multicolumn{1}{c}{$I^{G}(J^{P})$}&
\multicolumn{1}{c}{$Expt.(with\; summary\; T. \;status)$}&
\multicolumn{1}{c}{RF. \cite{go}}
 \\
\hline
\vspace{.1 cm}
$5$ & 3  & 90 & 0 & 1890 & $0^{+}(0^+)$ & $f_0(1590) : (included)$ & - \\
    &    &     &   &(1690)&              &                        &  \\
\vspace{.1 cm}
    &    &     & 2 & 2100 & $0^{+}(2^+)$ & $f_2(2150) : (omitted )$ & - \\
    &    &     &   &(1900)&              &                        &  \\
\hline
\vspace{.1 cm}
$6$ & 21  & 90 & 0 & 1890 & $0^{+}(0^+)$ &                        & 1990\\
    &    &     &   &(2090)&              &                        &     \\
\vspace{.1 cm}
    &    &     & 2 & 2100 & $0^{+}(2^+)$ &$f_2(2300) : (included)$ & 2240\\
    &    &     &   &(2300)&              &                        &  \\
\hline
\vspace{.1 cm}
$7$ & 21  & 81 & 0 & 1971 & $0^{+}(0^+)$ & $f_0(1590) : (included)$ & - \\
\vspace{.1 cm}
    &    &     & 2 & 2100 & $0^{+}(2^+)$ & $f_2(2010) : (included )$ & 2050\\
\hline
\vspace{.1 cm}
$8$ & 21  & 75 & 0 & 2125 & $0^{+}(0^+)$ &                          & - \\
\vspace{.1 cm}

    &    &     & 2 & 2335 & $0^{+}(2^+)$ & $f_2(2340) : (included )$ & - \\

\end{tabular}
\end{table}
\begin{table}
\centering
\caption{$\rho $-$\omega $ and $K^{*} $ families, $\alpha = 45\;
MeV$. All energies in $MeV$}

\begin{tabular}{cccccccc}
\\
\hline
\multicolumn{1}{c}{Band} &
\multicolumn{1}{c}{$[\nu ]$} &
\multicolumn{1}{c}{$C(\lambda , \mu )$} &
\multicolumn{1}{c}{ $L$ } &
\multicolumn{1}{c}{IBFM }&
\multicolumn{1}{c}{$I^{G}(J^{P})$}&
\multicolumn{1}{c}{$Expt.(with\; summary\; T. \;status)$}&
\multicolumn{1}{c}{RF. \cite{go}}
 \\
\hline
\vspace{.1 cm}
$1$ & 3  & 144 & 0 & 1420 & $1^{+}(1^-)$ & $\rho (1450) : (included)$ &1450 \\
    &    &     &   &      & $1/2(1^-)$ & $K^*(1410) : (included)$ & 1580 \\
    &    &     &   &      & $0^-(1^-)$ & $\omega (1390) : (included)$ &1630 \\
\vspace{.1 cm}
    &    &     & 2 & 1690 & $1^{+}(3^-)$ & $\rho _3(1690) : (included)$ &1680\\
    &    &     &   &      & $1/2(3^-)$ & $K^*(1780) : (included)$ & 1790 \\
    &    &     &   &      & $0^-(3^-)$ & $\omega 3(1670) : (included)$ &1680\\
\vspace{.1 cm}
    &    &     & 4 & 2320 & $1^{+}(5^-)$ & $\rho _5(2350) : (omitted)$ & 2300
\\
    &    &     &   &      & $1/2(5^-)$ & $K^{*}_5(2380) : (omitted )$ & 2390 \\
\vspace{.1 cm}
    &    &     & 1 & 1510 & $1/2(0^+)$ & $K^{*}_0(1430) : (included)$ & 1240 \\
    &    &     &   &      & $1/2(2^+)$ & $K^{*}_2(1430) : (included)$ & 1430 \\
\hline
\vspace{.1 cm}
$2$ & 21  & 114 & 0 & 1690 & $1^{+}(1^-)$ & $\rho (1700) : (included)$ &1660 \\
    &    &     &   &      & $1/2(1^-)$ & $K^*(1680) : (included)$ & 1780 \\
    &    &     &   &      & $0^-(1^-)$ & $\omega (1600) : (included)$ &1630 \\

    &    &     &   &      & $0^-(1^-)$ & $\phi (1680) : (included)$ &1660 \\
\vspace{.1 cm}
    &    &     & 2 & 1690 & $1^{+}(3^-)$ & $\rho _3(2250) : (omitted)$ &2130 \\
    &    &     &   &      & $1/2(3^-)$ & $K^{*}_3(1780) : (included)$ & - \\
    &    &     &   &      & $0^{-}(3^-)$ & $\phi_3(1850) : (included)$ &1900 \\
\vspace{.1 cm}
    &    &     & 3 & 2130 & $1/2(4^+)$ & $K^*_4(2045) : (included)$ & 2110 \\
\hline
\vspace{.1 cm}
$3$ & 111  & 108 & 1 & 1844 & $1/2(0^+)$ &$K^{*}_0(1950) : (omitted )$ & 1890
\\
    &    &     &     &      &  $1/2(2^+)$   &$K^*_2(1980) : (omitted)$ &1940\\
\hline
\vspace{.1 cm}
\end{tabular}
\end{table}

\newpage

\begin{thebibliography}{}
\bibitem{bar1}  A. O. Barut, Phys.Rev. B135 (1964) 839 ; Phys.
Lett. B26 (1968) 308.
\bibitem{bar2}  A. O. Barut, Phys. Rev. B139 (1965) 1107.
\bibitem{dot} Y. Dothan, M. Gell Mann and Y. Ne'eman, Phys.
Lett. 17 (1965) 283.
\bibitem{boh} {\it{Dynamical Groups and Spectrum Generating
Algebras}} eited by A. Bohm, Y. Ne'eman and A. O. Barut (World
Scientific 1989)
\bibitem{dk} C. De Tar and J. Kogut, Phys.Rev.Lett. 59 (1987) 399.
\bibitem{gott1} S. Gottlieb, W. Liu, D. Toussaint, R. L. Renken
and R. L. Sugar Phys.Rev.Lett.59 (1987) 1881.
\bibitem{kut} D. Kutasov and N. Seiberg, Nucl.Phys.B358 (1991) 600.
\bibitem{fr} P. Freund and J. Rosner, Phys.Rev.Lett.68 (1992) 765.
\bibitem{cud} J. Cudell and K. Dienes, Phys.Rev.Lett. 69 (1992) 1324.
\bibitem{ddt} J. Dey, M. Dey and L. Tomio, Phys. Lett. B 288
(1992) 306.
\bibitem{mus} M. Mustafa, A, Ansari, J. Dey and M. Dey, Phys.
Lett B 311 (1993) in press
\bibitem{cap} S. Capstick and N. Isgur, Phys. Rev. D 34 (1986) 2809
\bibitem{mur} M. Murthy, M. Dey, J. Dey and R. K. Bhaduri,
Phys.Rev.D 30 (1984) 152.
\bibitem{dey} J. Dey and M. Dey,  234 (1990) 349.
\bibitem{ell} J. P. Elliott, Proc. Roy. Soc. A 245 (1958) 128
\bibitem{kot} V. Kota, Phys. Rev. C 33 (1986) 2218.
\bibitem{yu}  Zu-Rong Yu, Mod. Phys. Lett. A 3 (1988) 1059.
\bibitem{kon} L. Kondratyuk, P. Volkovitsky and C. Guaraldo,
Phys. Lett. B 278 (1992) 475.
\bibitem{go} S. Godfrey,  and N. Isgur, Phys. Rev. D 32 (1985) 189.
\bibitem{sr} E. Shuryak and J. Rosner, Phys. Lett. B 218 (1989) 78
\bibitem{ir} F. Iachello, Phys. Rev. Lett. 62 (1989) 2440;
Comment by D. Robson and the answer, Phys. Rev. Lett. 63 (1989)
1890

\end{thebibliography}
\end{document}